\documentclass[12pt,aps,preprint,nofootinbib,superscriptaddress,nobalancelastpage]{revtex4}

\usepackage{hyperref}
\usepackage{epsfig}
\usepackage{amssymb}
\usepackage{amsmath}
\usepackage{amsfonts}
\usepackage{graphics}
\usepackage{cancel}
\usepackage{bbm}
\usepackage{mathrsfs}
\usepackage{slashed}
\usepackage{color}
\usepackage{axodraw}

\newcommand{\mtrx}[2]{\left(\begin{array}{#1} #2 \end{array}\right)}

\newcommand{\Ref}{Ref.}
\newcommand{\Sec}{Sec.}

\newcommand{\tab}{Tab.}

\newcommand{\fig}{Fig.}
\newcommand{\Fig}{Fig.}

\newcommand{\bea}{\begin{eqnarray}}
\newcommand{\eea}{\end{eqnarray}}

\newcommand{\be}{\begin{equation}}
\newcommand{\ee}{\end{equation}}

\newcommand{\ba}{\begin{array}}
\newcommand{\ea}{\end{array}}
\newcommand{\znbbeq}{0\nu\beta\beta }
\newcommand{\znbb}{$\znbbeq $}

\newcommand{\ie}{\emph{i.e.}}
\newcommand{\eg}{\emph{e.g.}}

\newcommand{\hc}{\mathrm{H.c.}}

\newcommand{\type}[1]{type-#1}

\allowdisplaybreaks

\begin{document}
\title{Heavy Neutrinos and Lepton Number Violation in $\boldsymbol{\ell p}$ Colliders}

\author{Carl Blaksley}
\email[]{blaksley@apc.univ-paris7.fr}
\affiliation{Max-Planck-Institut f\"ur Physik
(Werner-Heisenberg-Institut), F\"ohringer Ring 6, 80805 M\"unchen,
Germany}
\affiliation{Laboratoire Astroparticule et Cosmologie (APC), Universit\'e Paris 7/CNRS, 10 rue A. Domon et L. Duquet, 75205 Paris Cedex 13, France}

\author{Mattias Blennow}
\email[]{blennow@mppmu.mpg.de}
\affiliation{Max-Planck-Institut f\"ur Physik
(Werner-Heisenberg-Institut), F\"ohringer Ring 6, 80805 M\"unchen,
Germany}

\author{Florian Bonnet}
\email[]{florian.bonnet@pd.infn.it}
\affiliation{INFN Padova, Via Marzolo 8, I-35131, Padova, Italy}

\author{Pilar Coloma}
\email[]{p.coloma@uam.es}
\affiliation{Instituto de F\'isica Te\'orica, Universidad Aut\'onoma de Madrid,
28049 Madrid, Spain}

\author{\mbox{Enrique~Fernandez-Martinez}}
\email[]{enfmarti@cern.ch}
\affiliation{Max-Planck-Institut f\"ur Physik
(Werner-Heisenberg-Institut), F\"ohringer Ring 6, 80805 M\"unchen,
Germany}
\affiliation{CERN Physics Department, Theory Division, CH-1211 Geneva 23, Switzerland}

\begin{abstract}

We discuss the prospects of studying lepton number violating processes in order to identify Majorana neutrinos from low scale seesaw mechanisms at lepton-proton colliders. In particular, we consider the scenarios of colliding electrons with LHC energy protons and, motivated by the efforts towards the construction of a muon collider, the prospects of muon-proton collisions.  We find that present constraints on the mixing of the Majorana neutrinos still allow for a detectable signal at these kind of facilities given the smallness of the Standard Model background. We discuss possible cuts in order to further increase the signal over background ratio and the prospects of reconstructing the neutrino mass from the kinematics of the final state particles.

\end{abstract}



\preprint{CERN-PH-TH/2011-094, EURONU-WP6-11-32, IFT-UAM/CSIC-11-25, MPP-2011-50}

\maketitle

\section{Introduction}

The first light of the Large Hadron Collider (LHC) has marked the beginning of a new era of high-energy collider physics. This actualizes the concept of not only proton-proton collisions, but also what kind of physics could be probed in other high-energy collisions exploiting proton energies similar to the LHC ones. At the same time, it is the hope that new colliders will not only probe the mechanism behind electroweak symmetry breaking and the stabilization of the electroweak scale, but also trace and identify the source of other known shortcomings of the Standard Model (SM), such as the existence of neutrino masses and dark matter. In the case of neutrino masses, the most popular scenario is generally considered to be the type-I seesaw models~\cite{Minkowski:1977sc,Mohapatra:1979ia,Yanagida:1979as,GellMann:1980vs} and it has already been shown that low-scale realizations~\cite{Mohapatra:1986bd,Bernabeu:1987gr,Branco:1988ex,Buchmuller:1990du,Buchmuller:1991tu,Datta:1991mf,Ingelman:1993ve} of such models can give observable signals in collider experiments~\cite{Datta:1993nm,Almeida:2000pz,Panella:2001wq,delAguila:2005mf,delAguila:2005pf,Bray:2005wv,Han:2006ip,delAguila:2006dx,Atwood:2007zza,Bray:2007ru,deAlmeida:2007gc,delAguila:2007em,delAguila:2008cj}.

One of the most striking predictions of the type-I seesaw is that neutrinos are Majorana particles and thus, that lepton number is broken through their Majorana mass terms. One of the most promising probes of the Majorana nature of neutrinos is the search for neutrinoless double beta decay (\znbb), which would produce a distinct peak at the end of the double beta decay spectrum (for a recent review see \Ref~\cite{Bilenky:2010zz}). Any search for processes violating lepton number will typically have a signal suppressed by the tiny neutrino masses~\cite{Kersten:2007vk,Ibarra:2010xw}. While in low scale seesaw mechanisms this smallness is due to a cancellation of different large contributions, this very same cancellation will suppress any lepton number violating signal unless the energy probed in the process is such that some particular contributions are enhanced or suppressed due to kinematics. In \znbb{} searches, contributions from neutrinos with masses above the nuclear scale of $\sim 100$ MeV are suppressed with respect to those below (see \eg{} \Ref~\cite{Blennow:2010th} for a discussion) and the cancellation can be avoided. Here we will discuss the prospects of observing the manifestly lepton number violating processes $p\ell^\mp \to \ell^\pm jjj$. Similar to \znbb{}, the contributions of mass eigenstates above the energy probed by the colliders will be suppressed, evading the cancellation behind the small neutrino masses. Moreover, collider experiments, and proton-lepton colliders in particular can, unlike \znbb{} searches, produce these neutrinos on-shell via s-channel contributions with the consequent enhancement of the signal.

The rest of this paper is organized as follows: in \Sec~\ref{sec:low-scale}, we review the low-scale seesaw model and summarize the current bounds. We then go on to discuss in \Sec~\ref{sec:LNV}, within the low-scale seesaw model, the lepton number violating processes that constitute our signal. Section~\ref{sec:numerics} is dedicated to providing the details of our numerical computations, while in \Sec~\ref{sec:results} we present the results. Finally, in \Sec~\ref{sec:summary}, we summarize our results and give our conclusions.

\section{The low-scale seesaw model and present constraints}
\label{sec:low-scale}

Let us consider the Lagrangian of the standard \type{I} seesaw model, which consists of the  Standard Model (SM) Lagrangian plus an extra piece containing the allowed couplings between the SM fields and additional gauge singlet fermions (\ie, right-handed neutrinos) $N_\mathrm{R}^i$:
\begin{equation}\label{eq:The3FormsOfNuMassOp}
\mathscr{L} = \mathscr{L}_\mathrm{SM} - \left[\frac{1}{2} \overline{N_\mathrm{R}^i} M_{ij} N^{c j}_\mathrm{R} + (Y_{N})_{i\alpha}\overline{N_\mathrm{R}^i} \phi^\dagger
L^\alpha +\hc \right]\; .
\end{equation}
Here, $\phi$ denotes the SM Higgs field, which breaks the electroweak (EW) symmetry after acquiring its vacuum expectation value, $v_{\mathrm{EW}}$. In this work, we will use the basis in which $M$ is diagonal with real positive entries. After electroweak symmetry breaking, this produces a $6 \times 6$ mass matrix
\begin{equation}
	\mathcal M = \mtrx{cc}{0 & m_D^T \\ m_D & M},
\end{equation}
where $m_D = v_{EW} Y_N$, in the basis $[\nu^\alpha_L, (N_\mathrm{R}^{i})^c]^T$. In terms of the light mass eigenstates $\nu^i$ and the heavy eigenstates $N^i$, the neutrino flavor eigenstates can be written as 
\begin{equation}
  \nu^\alpha_L \simeq (\delta_{\alpha\beta}-\frac{1}{2} \theta_{\alpha j}\theta_{\beta j}^*)U_{\beta i} \nu^i + \theta_{\alpha i} N^i,
\end{equation}
where $\theta = m_D^\dagger M^{-1}$. In the cases where the heavy neutrinos can be integrated out, this implies a non-unitary mixing matrix $\mathcal N = (1 - \eta/2) U$, where $\eta = \theta \theta^\dagger$, for the light neutrinos~\cite{Langacker:1988ur,Nardi:1994iv,Tommasini:1995ii,Broncano:2002rw}. Inserting this relation into the electroweak interaction Lagrangian, we obtain the following coupling between the heavy neutrinos and the charged leptons via the $W$ bosons
\begin{equation}
	\mathscr L_{\rm int} = \frac{g}{\sqrt 2} W_\mu \overline{\ell^\alpha_L} \gamma^\mu \theta_{\alpha i} N^i + \hc.
\end{equation}

The unitarity deviations implied by the heavy neutrinos can thus be used to constrain the elements of $\theta$ through the relation $\eta = \theta \theta^\dagger$ which implies
\begin{equation}
	|\theta_{\alpha i}|^2 \leq |\eta_{\alpha\alpha}| = 0.005 \pm 0.005
\end{equation}
at the 90~\% Confidence Level (CL). This constraint stems from the effects that such terms would have in universality tests of the weak interactions as well as the invisble width of the $Z$~\cite{Antusch:2006vwa}. Notice that somewhat stronger bounds $|\eta_{e e}| < 4.0 \cdot 10^{-3}$, $|\eta_{\mu \mu}| < 1.6 \cdot 10^{-3}$ and $|\eta_{\tau \tau}| < 5.3 \cdot 10^{-3}$ are obtained when measurements of $G_F$ from muon decay and the CKM unitarity are assumed~\cite{Antusch:2008tz} instead of the invisible width of the $Z$. The combination of both data sets would, however, result in a larger allowed region since their respective best fits do not coincide. Indeed, while the invisible width of the $Z$ prefers non-zero values of $\eta$ in order to accommodate the present $2 \sigma$ deviation from the SM, such a result is not favored by the measuremets of $G_F$. For definiteness we will fix here the size of the signal by accommodating the data on the invisble width of the $Z$, setting  $|\eta_{\alpha \alpha}| = 5.0 \cdot 10^{-3}$. However, the rescaling of the signal under study to smaller mixing angles is trivial through an overall quadratic dependence, as we will discuss later.

If the masses of the heavy neutrinos display a moderate hierarchy, the signal will mainly be dominated by the contribution of the mass eigenstate within the reach of the collider energy. Thus, in order to simplify the discussion, we will assume there is only one heavy neutrino contributing to the signal, while the other neutrinos, necessary for the generation of the observed pattern of masses and mixings at low energies, are heavier and do not contribute significantly. The mixing of this neutrino with the SM fermions will be set to $ |\theta_{\alpha 1}|^2 \equiv |\theta_{\alpha}|^2 = 0.005 $ as discussed above.

Note that, if we also impose the condition that neutrino masses are small while maintaining large Yukawa couplings, then an additional constraint $|\eta_{\alpha\alpha}|\cdot|\eta_{\beta\beta}| \simeq |\eta_{\alpha\beta}|^2$ applies~\cite{Buchmuller:1990du,Datta:1991mf,Ohlsson:2010ca}. In particular, strong constraints on the product will then stem from processes such as $\mu \to e\gamma$. Therefore, we will consider scenarios with only one $\theta_{\alpha} \neq 0$, where the mixing is to only one flavour. Thus, we will not consider lepton flavour violation in the signal but only lepton number violation.

%The mixing of this neutrino with the SM fermions is thus bounded by $ |\theta_{\alpha i}|^2 \leq 0.005 $. Moreover, as stated above, we will assume that this mixing is to only one flavour. Thus we will not consider lepton flavour violation in the signal but only lepton number violation.

Finally, it should be noted that \znbb{} decay searches in general provide strong bounds on $\theta_{ei}$ \cite{Ibarra:2010xw}. However, it is possible to satisfy both the constraints from \znbb{} and bounds on neutrino masses while still maintaining large Yukawa couplings~\cite{Ingelman:1993ve}.

\section{Lepton number violating processes}
\label{sec:LNV}

%\subsection{Signal}

Here we will consider the process $p \ell^- \to \ell^+ jjj$ to investigate the possibility of producing the heavy right-handed neutrinos of the type-I Seesaw mechanism and observe the lepton-number-violating signals associated to their Majorana nature at colliders. The advantage of lepton-proton collisions with respect to purely hadron or lepton colliders is the cleanness of the signal. Indeed, since the lepton number in the initial state is one, the observation of a charged antilepton in the final state is a clear signal of violation of lepton number by two units, as long as there is no missing energy. This last requisite translates in the presence of three jets in the final state, two of them reconstructing an invariant mass equal to the $W$ mass. Conversely, lepton number violation in hadronic or $\ell^- \ell^+$ colliders implies tagging two leptons of the same charge in the final state together with either the presence of missing energy in the form of neutrinos or a higher number of jets, making the signal more challenging to search for. 

\begin{figure}
  \begin{center}
    %% Non-mixing
     \begin{picture}(220,140)(0,0)
       %\BBox(0,0)(220,140)
       \ArrowLine(10,10)(60,10)
       \Text(8,10)[r]{$u$}
       \ArrowLine(60,10)(210,10)
       \Text(212,10)[l]{$j$}
       \ArrowLine(210,40)(160,70)
       \Text(212,42)[lt]{$j$}
       \ArrowLine(160,70)(210,90)
       \Text(212,88)[lb]{$j$}
       \Photon(160,70)(110,90){2}{5.5}
       \Text(133,78)[tr]{$W$}
       \ArrowLine(210,130)(110,90)
       \Text(212,130)[lb]{$\ell^+$}
       \ArrowLine(10,90)(60,90)
       \Text(8,90)[r]{$\ell^-$}
       \Line(60,90)(110,90)
       \Text(85,92)[b]{$N_R$}
       \Photon(60,90)(60,10){-2}{6.5}
       \Text(64,50)[l]{$W$}
     \end{picture}
     \hspace{0.02\textwidth}
    %% Mixing
     \begin{picture}(170,140)(0,0)
       %\BBox(0,0)(170,140)
       \ArrowLine(10,10)(60,10)
       \Text(8,10)[r]{$u$}
       \ArrowLine(10,90)(60,90)
       \Text(8,90)[r]{$\ell^-$}
       \Line(60,90)(110,50)
       \Text(87,72)[lb]{$N_R$}
       \Photon(110,50)(60,10){2}{5}
       \Text(83,32)[rb]{$W$}
       \Photon(60,90)(110,110){2}{4.5}
       \Text(83,102)[rb]{$W$}
       \ArrowLine(110,110)(160,130)
       \Text(162,128)[bl]{$j$}
       \ArrowLine(160,90)(110,110)
       \Text(162,92)[tl]{$j$}
       \ArrowLine(160,50)(110,50)
       \Text(162,50)[l]{$\ell^+$}
       \ArrowLine(60,10)(160,10)
       \Text(162,10)[l]{$j$}
     \end{picture}
    \caption{The Feynman diagrams contributing to the process $\ell^-u \to \ell^+jjj$ at tree level in the low-scale seesaw. The left diagram includes a right-handed neutrino in an $s$-channel, while the right diagram has the right-handed neutrino in a $t$-channel. The left diagram is enhanced when the right-handed neutrino can be produced on shell.}\label{fig:diagrams}
  \end{center}
\end{figure}
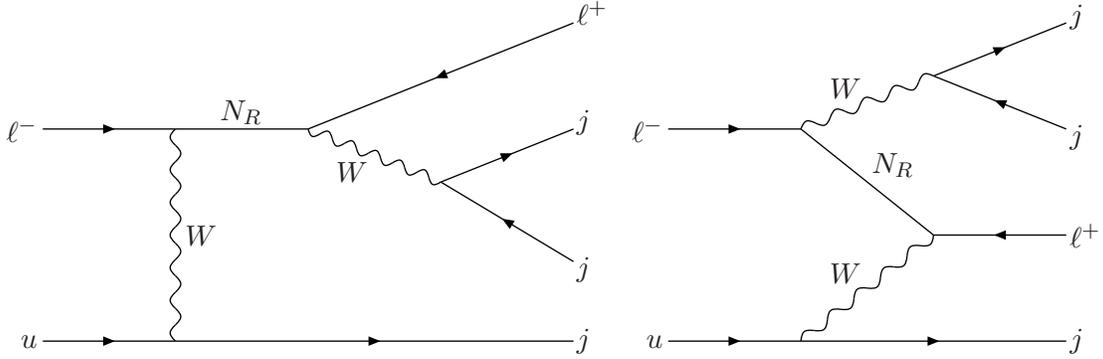

In the \type{I} seesaw, two diagrams contribute to the signal under study (see \Fig~\ref{fig:diagrams}).
The diagram with the Majorana neutrino in $s$-channel will be enhanced and dominate the signal if the collider energy is high enough to produce it on shell. We will therefore explore this possibility, which can help overcome the expected suppression of the signal by the smallness of neutrino masses. Thus, the process basically corresponds to the on shell production of the heavy neutrino, via the exchange of a $W$ boson, and its subsequent decay to $\ell^+ W^-$ with the $W$ decaying hadronically. The branching ratio for the first decay is roughly 25~\%~\cite{Buchmuller:1990vh}. On the other hand, the production process is weighted by the mixing $|\theta_{\alpha}|^2$ of the heavy neutrino $N$ with the charged lepton $\ell_\alpha$. Thus, it is trivial to extend the analysis from the assumed values of the mixing angles of $|\theta_{\alpha}|^2 = 5 \cdot 10^{-3}$ to smaller mixings by a general rescaling.

The electron-proton process we discuss here was already studied in \cite{Buchmuller:1990vh,Buchmuller:1991tu,Buchmuller:1992wm,Ingelman:1993ve}, mainly focusing on the DESY experiment but also extended to a combination of LEP and LHC. The production cross section and the decay rate of the heavy neutrino were computed in order to estimate the rate of the signal that could be expected in a electron-proton collider. In \Ref~\cite{Ingelman:1993ve} a first numerical simulation of the signal was also performed. The analysis presented here is motivated by the improved simulation techniques presently available, in particular regarding hadronization processes, and better knowledge of the parton distribution functions (PDFs) of the proton. We also extend their analysis to the higher collider energies discussed in the LHCeC proposal~\cite{LHCeC}, which results in an enhancement of the signal and sensitivity to higher neutrino masses. We have therefore extended the original analysis to a wider range of neutrino masses than initially considered and showed explicitly the dependence of the signal on the mediating neutrino mass. 

Furthermore, we also explore for the first time more sophisticated collider technologies, such as muon-proton colliders, inspired by the recent efforts towards a muon collider (see \eg{} \Ref~\cite{Shiltsev:2010qg} for a recent status review). This setup would allow to explore different matrix elements with respect to the electron-proton option, providing a completely independent search channel also with respect to \znbb{} searches. Moreover, the energies that could be expected from such a facility imply a huge gain in sensitivity allowing to explore much smaller mixings or larger sterile neutrino masses than their electron counterparts. In both scenarios we will assume a detector setup based on the capabilities of ATLAS and CMS.

%\subsection{Background}
Regarding possible sources of background, the manifest lepton number violating nature of the signal in absence of missing energy makes it very clean and difficult to mimic by SM processes. Several sources of background were already discussed in \Ref~\cite{Ingelman:1993ve} where it was found that, for the lepton number violating signal under study, the dominant background stems from $W$ production with its subsequent decay into the $\ell^+$ required for the signal\footnote{Other sources of background discussed in \Ref~\cite{Ingelman:1993ve} such as boson-gluon fusion and DIS are negligible because of the lepton number violating nature of the signal considered.}. The original $\ell^-$ can be missed for example if it becomes a neutrino or is lost in the beamline. We have simulated and studied these two sources of background. In the second case the contribution to $W$ production is dominated by the exchange of an almost real photon with a very collinear outgoing electron. In order to compute this process, we have simulated a proton-photon collision and convoluted it with the photon PDF in a charged lepton. This turns out to be the dominant source of background, in particular when the two final state jets are produced by QCD processes. In \Ref~\cite{Ingelman:1993ve}, this background was reduced by means of a cut in the invariant mass of the two jets in the final state $m_{jj}$ compatible with that of a $W$ boson $M_W$, complemented with a cut in the minimum transverse momentum of the outgoing $\ell^+$ ($p_{T,\ell^+}$). However, as we will show in Sect.~\ref{sec:results}, the presence of neutrinos in the final state, necessary to produce the final state positron through SM processes, suggests also the use of cuts in missing transverse energy ($E_{T,\rm miss}$) to suppress the background. We have found that cuts in maximum $E_{T,\rm miss}$ and minimum $p_{T,\ell^+}$ actually provide a better signal/background ratio than those in $m_{jj}$.
Moreover, in the case of the $p \ell^- \to \ell^+ jjj \nu_l \nu_l$ background, two of the jets typically originate from a $W$ boson decay. Thus, the invariant mass cut around $M_W$ will not avail to reduce this source of background. On the other hand, the cuts in maximum $E_{T,\rm miss}$ and minimum $p_{T,\ell^+}$ prove to be useful for this other background too. This extra cut in $E_{T,\rm miss}$ thus represents a very useful complementary tool to the cut in $p_{T,\ell^+}$ in order to reduce all sources of SM background.    

\section{Numerical analysis}
\label{sec:numerics}

We base our numerical study of the signal and background on the use of the MadGraph/MadAnalysis~\cite{Alwall:2007st} software tools supplemented by Pythia~\cite{Sjostrand:2006za} and PGS~\cite{PGS} to process the resulting events. The type-I seesaw description for MadGraph/MadAnalysis was obtained via the FeynRules~\cite{Christensen:2008py} software. For the protons we assume an LHC-like beam with an energy of 7~TeV, while several different choices are examined for the lepton beam. In the case of an electron beam we study both the conservative setup with a beam energy of 50~GeV and a more optimistic setup where the beam energy is 150~GeV~\cite{LHCeC}. For muons, on the other hand, we consider the beam energies that have been discussed for a future muon collider. These vary between 500~GeV for the more conservative case and 2~TeV for the more optimistic proposals~\cite{Shiltsev:2010qg}. In all cases under study, we employ a detector with capabilities similar to the ATLAS and CMS detectors.
When generating events, we impose the acceptance cuts specified in \tab~\ref{tab:gcuts}.
\begin{table}
\begin{center}
\begin{tabular}{|lccc|}
\hline
{\bf Variable} & {\bf Jets} & {\bf Leptons} & {\bf Photons} \\
\hline
$\; p_T$ & $> 20$~GeV & $> 10$~GeV & $> 10$~GeV \\
$\; \eta$   & $< 5$      & $< 2.5$    & $< 2.5$ \\
\hline
\end{tabular}
\caption{Acceptance cuts used in MadGraph for our simulations.}\label{tab:gcuts}
\end{center}
\end{table}

For the simulation of the SM background process $p \ell^- \to \ell^+ j j j \nu \nu$, we use the default SM implementation provided by the MadGraph distribution. In the case of the $p \ell^- \to \ell^- \ell^+ j j j \nu $, which is dominated by the exchange of an almost real photon with a very collinear outgoing electron, we simulate instead the process $p \gamma \to \ell^+ j j j \nu $ convoluted with the PDF of a photon inside the charged lepton.

For each experimental setup, we simulate the signal for the case of a heavy neutrino mass of $M_N = 250$, 500, and 750~GeV while setting the mixing to the value required to accommodate the invisible width of the $Z$: $|\theta_{\alpha}|^2 = 0.005$. This provides a quite optimistic signal, but the scaling of the results is trivial with $|\theta_{\alpha}|^2$, as previously discussed. For all cases, we generate a total of $8 \cdot 10^4$ events satisfying the acceptance cuts. At a later stage, after presenting the bare results of the simulation, we will implement additional cuts to reduce the background. The implementation of these cuts will be made at the MadAnalysis level.

\section{Results}
\label{sec:results}

As we are discussing new physics searches, it is fundamental to distinguish between the signal and the SM background. Thus, in order to increase the sensitivity of these searches, cuts which can suppress the background while not affecting the signal too adversely should be investigated. Naturally, an expected background lower than one event would not be very worrisome and the constraining factor would then be one of the signal cross section. However, as the achievable luminosities for these speculative facilities are uncertain, we will present cuts designed to enhance the signal-to-background ratio regardless of the smallness of the cross section. For definiteness, we will assume a baseline integrated luminosity of $\sim100$~fb$^{-1}$ and include this value in the plots for reference. Indeed, this value is in the ballpark of the $10^{33}$~cm$^{-2}$s$^{-1}$ being discussed for the LHCeC proposal~\cite{LHCeC} for its lower energy version. For a 150~GeV electron beam the luminosity could be around an order of magnitude smaller. As for a prospective muon-proton collider the luminosities under discussion for a muon collider range between $10^{33}$~cm$^{-2}$s$^{-1}$ and $10^{35}$~cm$^{-2}$s$^{-1}$~\cite{Geer:2009zz}.

The major question is thus what separates the signal events from the background ones. A cut that has been previously studied in the literature is to reject events where the outgoing $\ell^+$ does not have a minimum transverse momentum. The reason for this cut to reject the background more strongly than the signal is that, while the signal is mainly due to on-shell heavy neutrinos, the background kinematics are such that the positron is predominantly emitted in the beam direction. Since all of the SM backgrounds include at least one neutrino, which will constitute missing energy, while the signal only includes particles visible to the detector, another viable cut could be to impose an upper bound on the missing transverse energy of the accepted events.

\begin{figure}
\begin{center}
\includegraphics[width=0.48\textwidth]{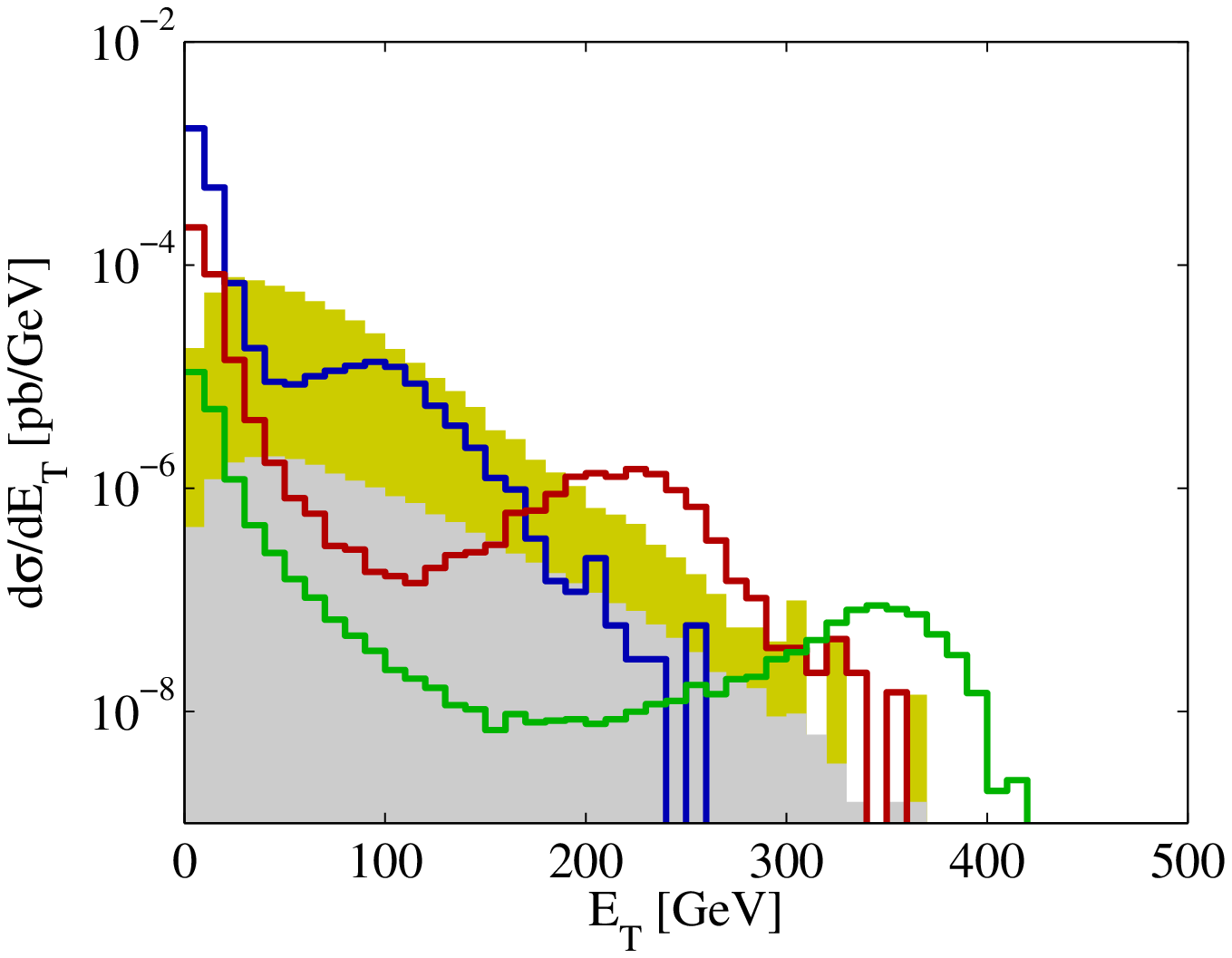}
\includegraphics[width=0.48\textwidth]{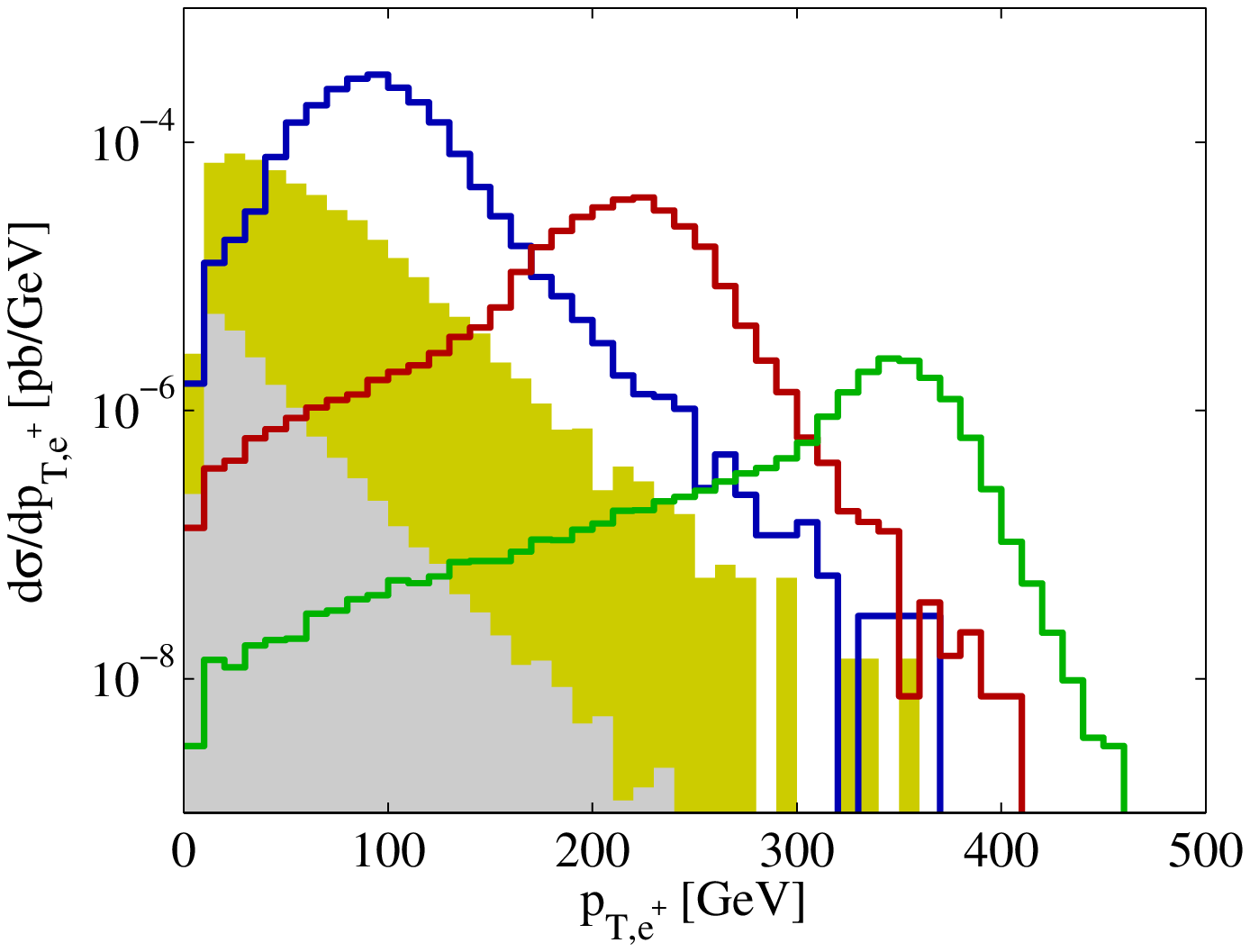}
\caption{Results for the differential cross sections for the 50~GeV electron beam facility. Here, $E_T$ is the missing transverse energy and $p_{T,e^+}$ is the transverse momentum of the positron. Blue, red, and green lines lines correspond to the signal simulated for heavy neutrino masses of $M_N = 250$, 500 and 750~GeV, respectively. The shaded area shows the main backgrounds, the smaller $p e^- \to e^+ j j j \nu_e \nu_e$ in light gray and the larger $p \gamma \to e^+ j j j \nu_e$ in green.}
\label{fig:nocut_e50}
\end{center}
\end{figure}

\begin{figure}
\begin{center}
\includegraphics[width=.48\textwidth]{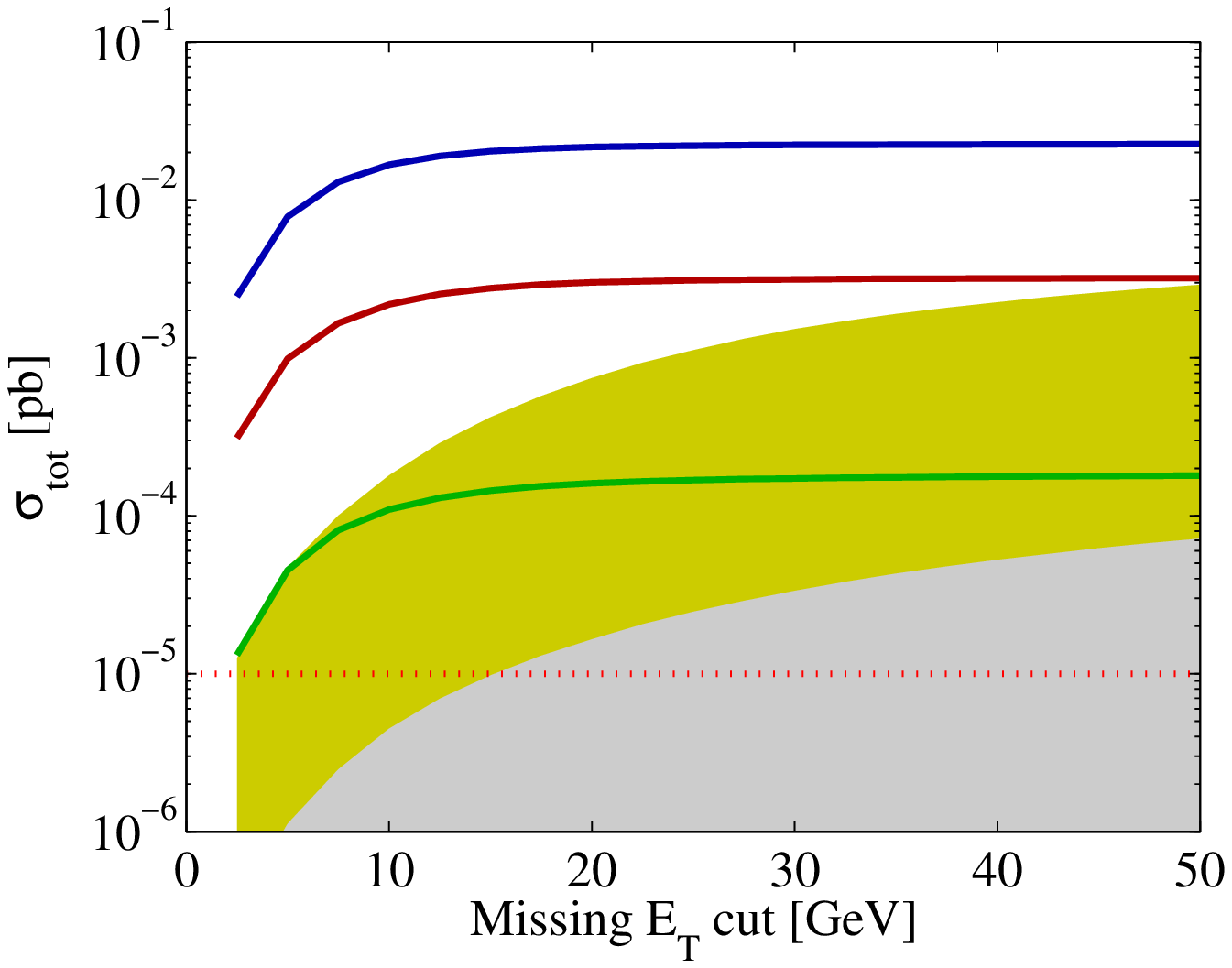}
\includegraphics[width=.48\textwidth]{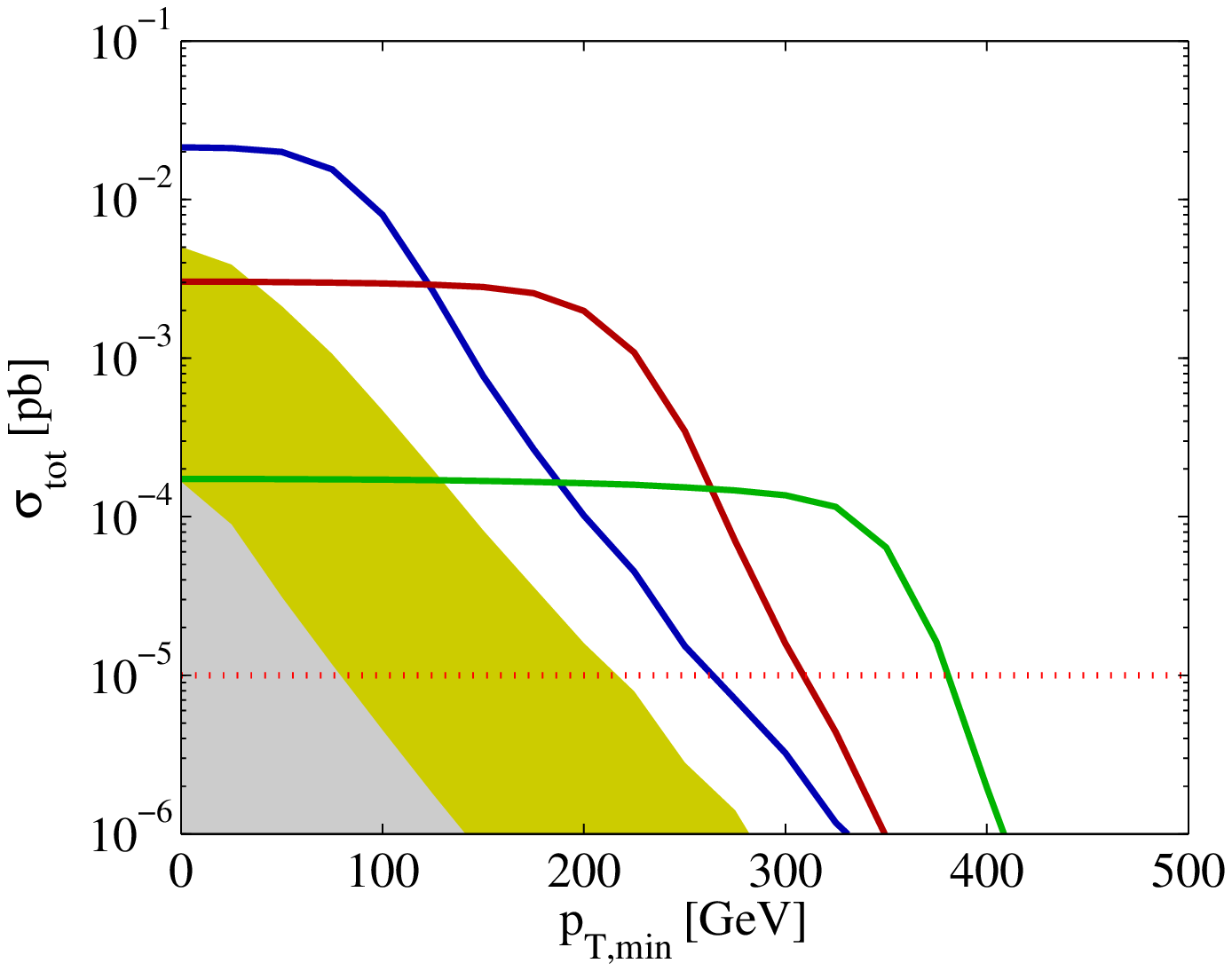} \\
\includegraphics[width=.48\textwidth]{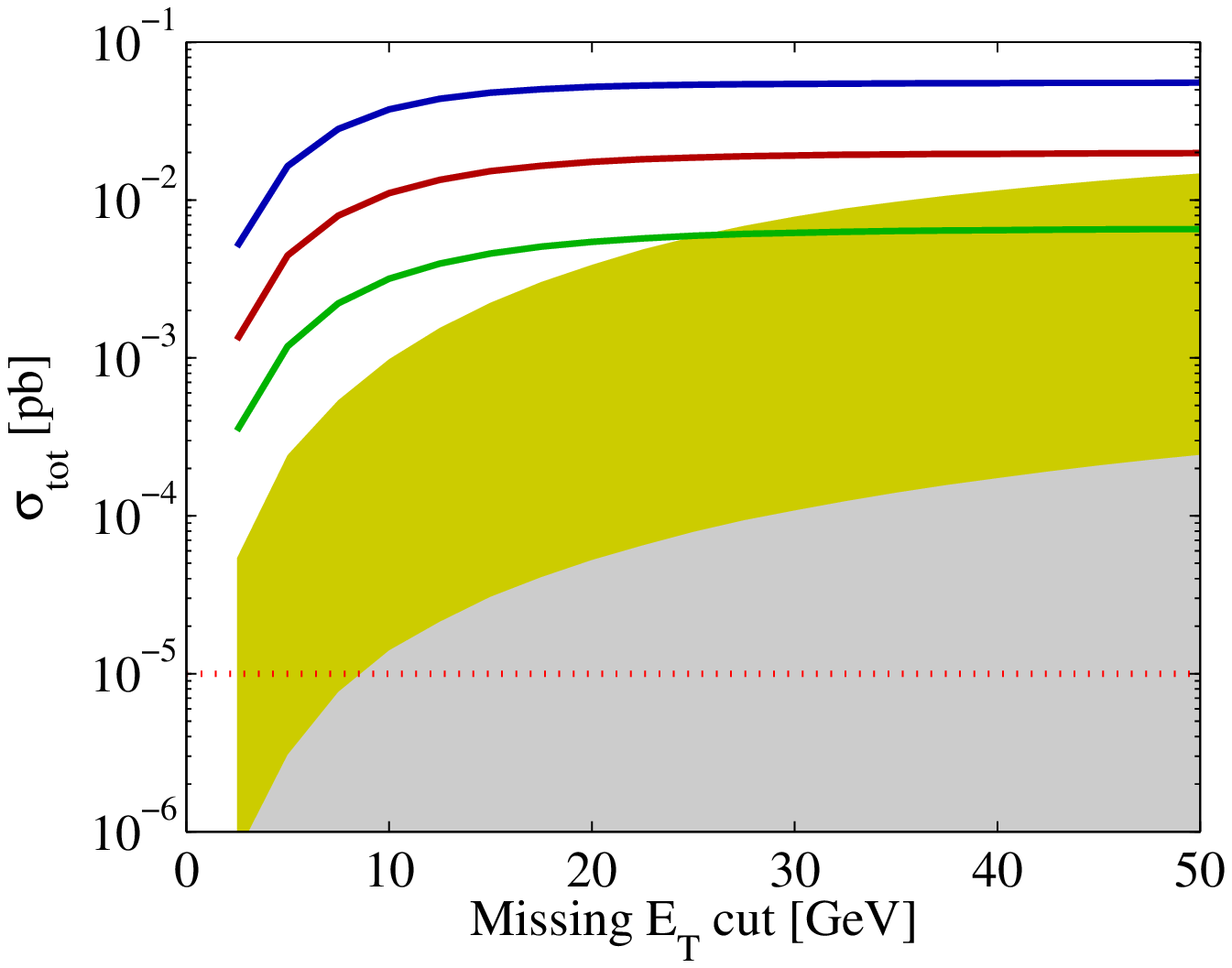}
\includegraphics[width=.48\textwidth]{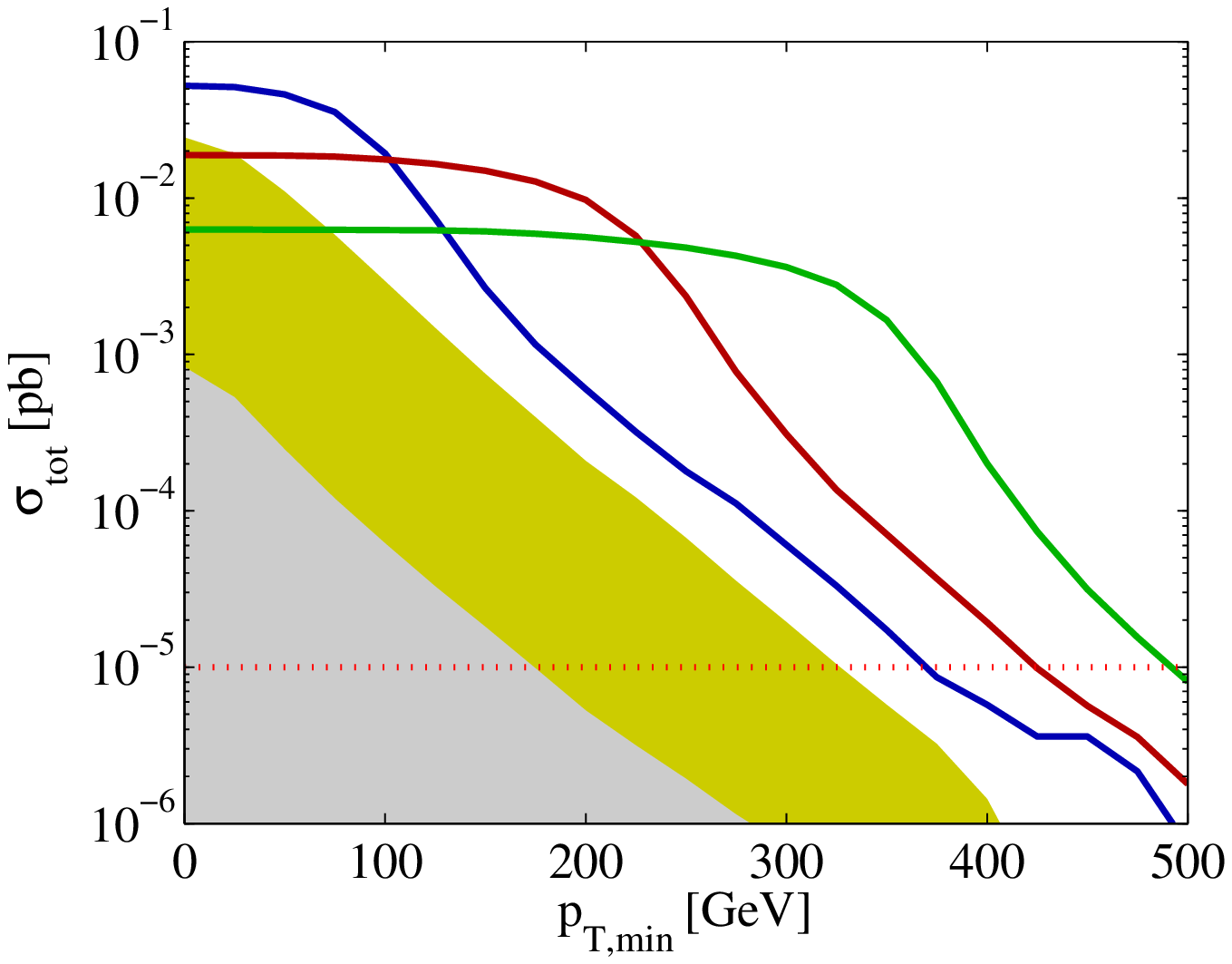}
\caption{Total cross section for the 50~GeV (upper) and 150~GeV (lower) electron beam facilities as a function of a cut on the maximum missing transverse energy (left) and the minimum transverse momentum of the positron (right). The blue, red, and green lines lines correspond to simulated heavy neutrino masses of $M_N = 250$, 500 and 750~GeV, respectively.  The shaded area shows the main backgrounds, the smaller $p e^- \to e^+ j j j \nu_e \nu_e$ in light gray and the larger $p \gamma \to e^+ j j j \nu_e$ in green. The horizontal line represents a cross section of $10^{-2}$~fb, which is the required cross section to have one expected event at an integrated luminosity of 100~fb$^{-1}$.}
\label{fig:cutse}
\end{center}
\end{figure}

In \fig~\ref{fig:nocut_e50}, we show our numerical results for the signal and background distributions in both of these variables at an electron beam energy of 50~GeV in order to illustrate how the signal and background differ.
The behavior for different beam energies, and for the muon beams, is similar. In order to illustrate the impact a given cut would have, we show in \fig~\ref{fig:cutse} how the signal and backgrounds for an electron beam would be affected as a function of the value at which the cut is implemented.
As can be seen from these figures, both cuts are indeed effective in increasing the signal-to-background ratio, although there are some qualitative differences between them. In particular, due to the peaked nature of the signal in the transverse momentum of the outgoing lepton, the cut on this variable is starting to deteriorate also the signal quite early in the cases of low $M_N$. On the other hand, the shape of the transverse energy distribution is similar for all $M_N$ in the area close to zero missing transverse energy, thus leading to a uniform behavior for the signal suppression. It should also be noted that even a very mild cut on the missing transverse energy would be enough to reduce the background by a factor of a few without noticeably affecting the signal and thus, at least a modest cut should be implemented in any analysis. For illustration, we will implement cuts of $p_{T,\ell^+} > 70$~GeV and $E_{T,\rm miss} < 10$~GeV. This roughly corresponds to the regions where the signals start to be noticeably reduced. 

The impact of the different cuts can be seen from the remaining total cross sections, which are presented in \tab~\ref{tab:xsects}. In the table we also show the cut in the jets invariant mass $m_{jj}$ around the mass of the $W$ performed in \Ref~\cite{Ingelman:1993ve} for comparison. It can be seen that this cut strongly reduces the signal while similar backgrounds suppressions can be achieved through the cuts in  $p_{T,\ell^+}$ and $E_{T,\rm miss}$ instead.
\begin{table}
\begin{center}
\scalebox{0.9}{
\begin{tabular}{cc|c|c|c|c}
Cuts & $M_n$ [GeV] & $ep$ [50~GeV]  & $ep$ [150~GeV]   & $\mu p$ [500~GeV] & $\mu p$ [2~TeV] \\
\hline
No cuts 
  & 250  & $23.53 \pm 0.04$  & $57.53 \pm 0.09$  & $112.3 \pm 0.2$  & $131.9 \pm 0.2$  \\
  & 500  & $3.351 \pm 0.006$  & $20.81 \pm 0.04$  & $56.32 \pm 0.11$  & $113.1 \pm 0.2$  \\
  & 750  & $0.1924 \pm 0.0010$  & $6.979 \pm 0.020$  & $30.35 \pm 0.07$  & $76.49 \pm 0.19$  \\
  & BG$2\nu$   & $0.1884 \pm 0.0004$  & $0.9478 \pm 0.0020$  & $3.554 \pm 0.014$  & $11.08 \pm 0.032$  \\
  & BG$1\nu$   & $5.576 \pm 0.012$  & $27.13 \pm 0.08$  & $100.1 \pm 0.3$  & $284.8 \pm 2.2$  \\
\hline
$p_{T,\ell^+} > 70$~GeV 
  & 250  & $16.65 \pm 0.05$  & $38.20 \pm 0.12$  & $80.1 \pm 0.5$  & $73.20 \pm 0.26$  \\
  & 500  & $2.994 \pm 0.007$  & $18.6 \pm 0.04$  & $51.41 \pm 0.11$  & $99.49 \pm 0.19$  \\
  & 750  & $0.1721 \pm 0.0009$  & $6.272 \pm 0.020$  & $28.33 \pm 0.07$  & $70.16 \pm 0.20$  \\
  & BG$2\nu$   & $0.0142 \pm 0.0002$  & $0.1381 \pm 0.0013$  & $0.8003 \pm 0.0061$  & $3.366 \pm 0.021$  \\
  & BG$1\nu$   & $1.227 \pm 0.012$  & $6.65 \pm 0.04$  & $29.07 \pm 0.16$  & $95.1 \pm 3.5$  \\
\hline
$E_{T,\rm miss} < 10$~GeV 
  & 250  & $16.71 \pm 0.05$  & $37.48 \pm 0.11$  & $77.1 \pm 0.5$  & $88.46 \pm 0.26$  \\
  & 500  & $2.182 \pm 0.008$  & $11.07 \pm 0.04$  & $32.39 \pm 0.11$  & $72.69 \pm 0.21$  \\
  & 750  & $0.1096 \pm 0.0007$  & $3.177 \pm 0.016$  & $15.09 \pm 0.06$  & $43.90 \pm 0.20$  \\
  & BG$2\nu$   & $0.0045 \pm 0.0001$  & $0.0140 \pm 0.0004$  & $0.0373 \pm 0.0013$  & $0.0691 \pm 0.0031$  \\
  & BG$1\nu$   & $0.180 \pm 0.005$  & $0.981 \pm 0.019$  & $3.33 \pm 0.06$  & $11.1 \pm 1.4$  \\
\hline
$p_{T,\ell^+} > 70$~GeV 
  & 250  & $12.29 \pm 0.04$  & $25.62 \pm 0.10$  & $57.0 \pm 0.6$  & $54.80 \pm 0.24$  \\
$E_{T,\rm miss} < 10$~GeV
  & 500  & $2.018 \pm 0.008$  & $10.24 \pm 0.04$  & $29.92 \pm 0.11$  & $66.98 \pm 0.21$  \\
  & 750  & $0.1019 \pm 0.0006$  & $2.97 \pm 0.016$  & $14.36 \pm 0.06$  & $41.40 \pm 0.20$  \\
  & BG$2\nu$   & $0.0002 \pm 0.0001$  & $0.0008 \pm 0.0001$  & $0.0038 \pm 0.0004$  & $0.0090 \pm 0.0011$  \\
  & BG$1\nu$   & $0.1006 \pm 0.0037$  & $0.546 \pm 0.014$  & $2.05 \pm 0.05$  & $8.2 \pm 1.3$  \\
\hline
$75$~GeV~$< m_{jj}$
  & 250  & $1.734 \pm 0.021$  & $3.07 \pm 0.05$  & $5.670 \pm 0.2460$  & $4.960 \pm 0.089$  \\
$m_{jj} < 85$~GeV
  & 500  & $0.995 \pm 0.008$  & $3.838 \pm 0.031$  & $9.68 \pm 0.08$  & $20.80 \pm 0.15$  \\
  & 750  & $0.0457 \pm 0.0004$  & $1.377 \pm 0.011$  & $5.07 \pm 0.05$  & $12.67 \pm 0.13$  \\
  & BG$2\nu$   & $0.0128 \pm 0.0002$  & $0.0475 \pm 0.0008$  & $0.1354 \pm 0.0025$  & $0.385 \pm 0.008$  \\
  & BG$1\nu$    & $0.341 \pm 0.007$  & $1.545 \pm 0.022$  & $5.01 \pm 0.08$  & $13.7 \pm 1.6$  \\\end{tabular}}
\caption{The total cross sections in units of fb after applied cuts for the different facilities discussed in the text.}
\label{tab:xsects}
\end{center}
\end{table}
As can be seen from the table the two cuts are fairly independent and complementary in nature, since their combination provides a significant background reduction with respect to the implementation of only one of them. Moreover, depending on the facility, luminosity and part of the parameter space that is being explored, some combination of the two would provide an optimal cut. %It should be noted that both cuts reduce the background to a level of $\mathcal O(10^{-5})$~fb and, therefore, to an in principle negligible level with the baseline assumption of an integrated luminosity of 100~fb$^{-1}$.

\begin{figure}
\begin{center}
\includegraphics[width=0.48\textwidth]{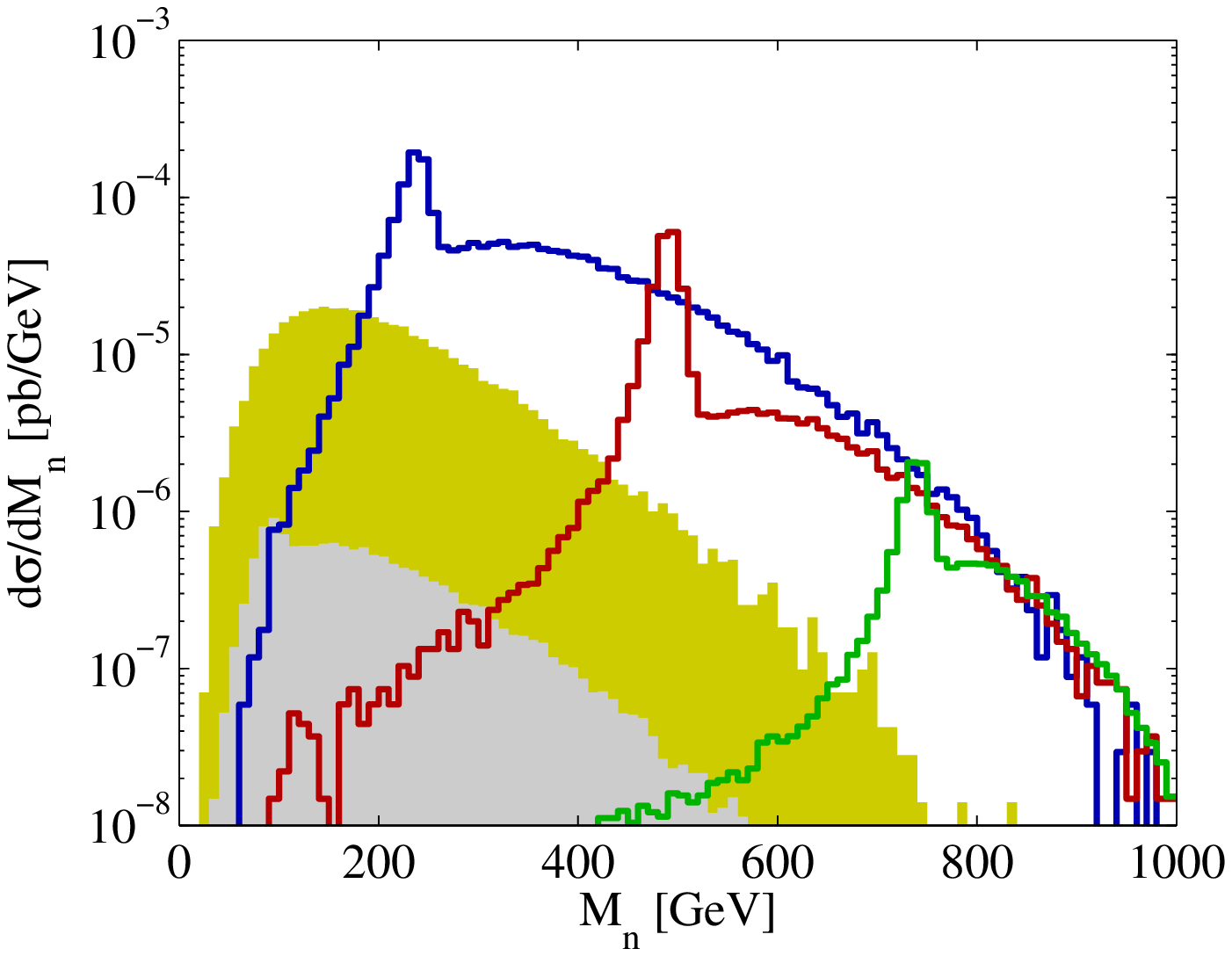}
\includegraphics[width=0.48\textwidth]{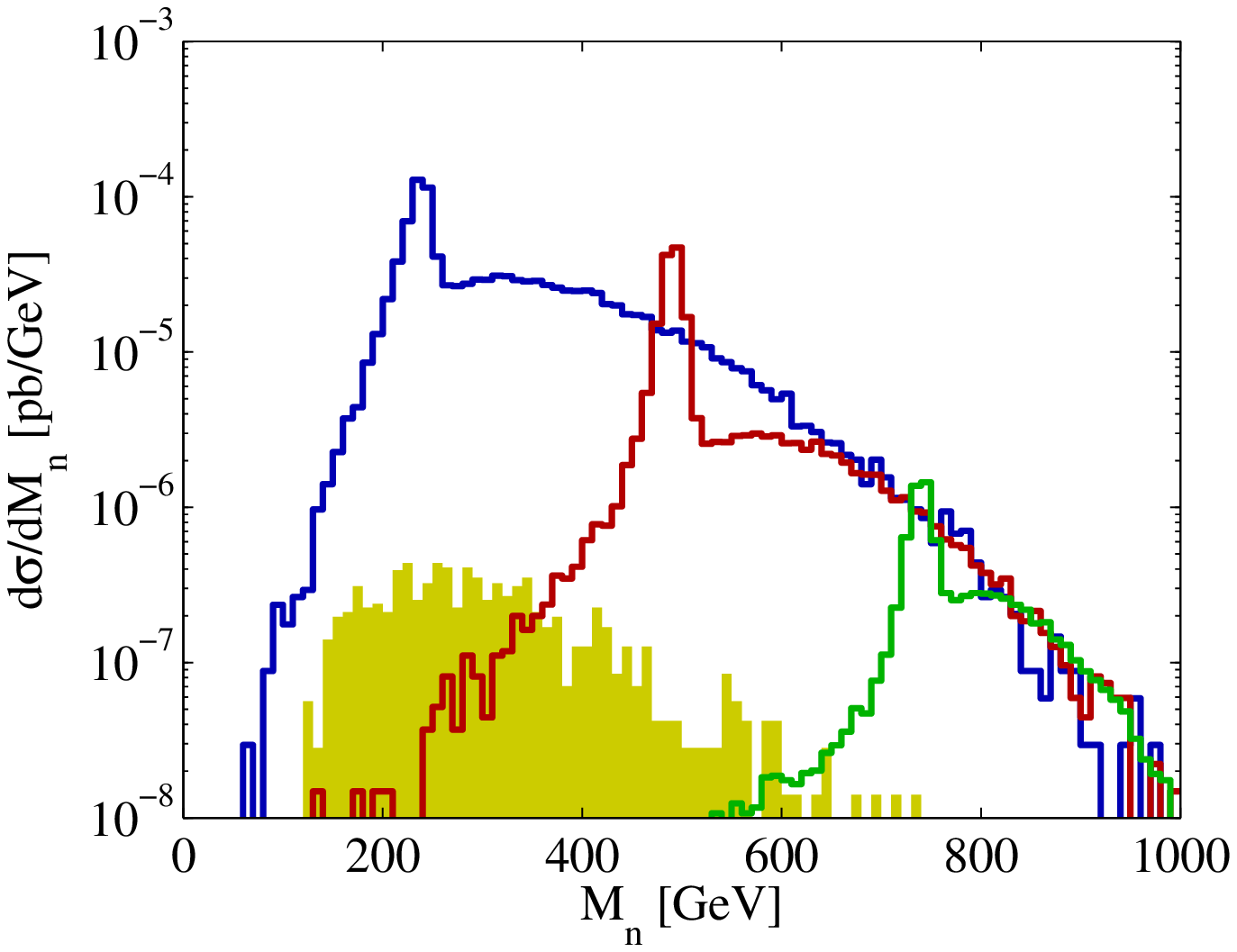} \\
\caption{The differential cross section with respect to the reconstructed mass $M_{jj\ell}$ at the 50~GeV electron beam facility. The results are shown both without (left) and with (right) cuts on $p_{T,e^+}$ and $E_{T,\rm miss}$. Blue, red and green lines correspond to the signals for heavy neutrino masses of $M_N = 250$, 500 and 750~GeV, respectively.  The shaded area shows the main backgrounds, the smaller $p e^- \to e^+ j j j \nu_e \nu_e$ in light gray and the larger $p \gamma \to e^+ j j j \nu_e$ in green.}
\label{fig:Mnjje}
\end{center}
\end{figure}

Another interesting kinematic variable is the reconstructed invariant mass of two jets and the outgoing lepton $M_{jj\ell}$. In the case of the signal, this should provide a measurement of the mass of the on-shell intermediate heavy neutrino, and thus, the majority of all signal events cluster around this value. A bump of events at a certain value would not only be in favor of signal over background, but also indicate the actual mass of the intermediate particle. Even if a search of this sort is performed, the cuts previously discussed still provide a complementary suppression of the background. This is illustrated in \fig~\ref{fig:Mnjje}, where we show the differential cross section with respect to $M_{jj\ell}$ at the 50~GeV electron-proton collider as an example.
An additional cut (not discussed in detail here) which could in principle be exploited to enhance the signal over possible backgrounds is the isolation of the final state antilepton. 

\begin{figure}
\begin{center}
\includegraphics[width=0.48\textwidth]{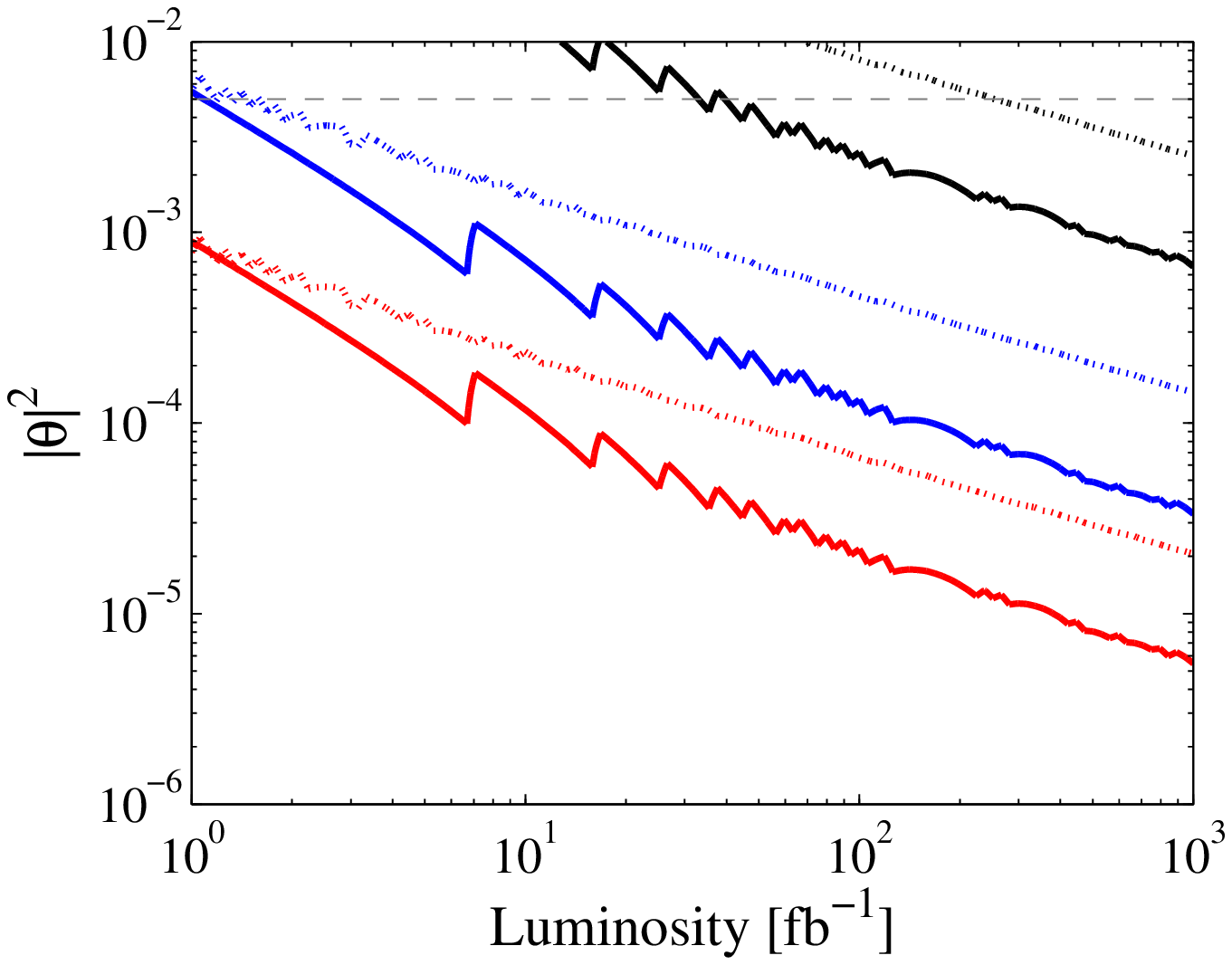}
\includegraphics[width=0.48\textwidth]{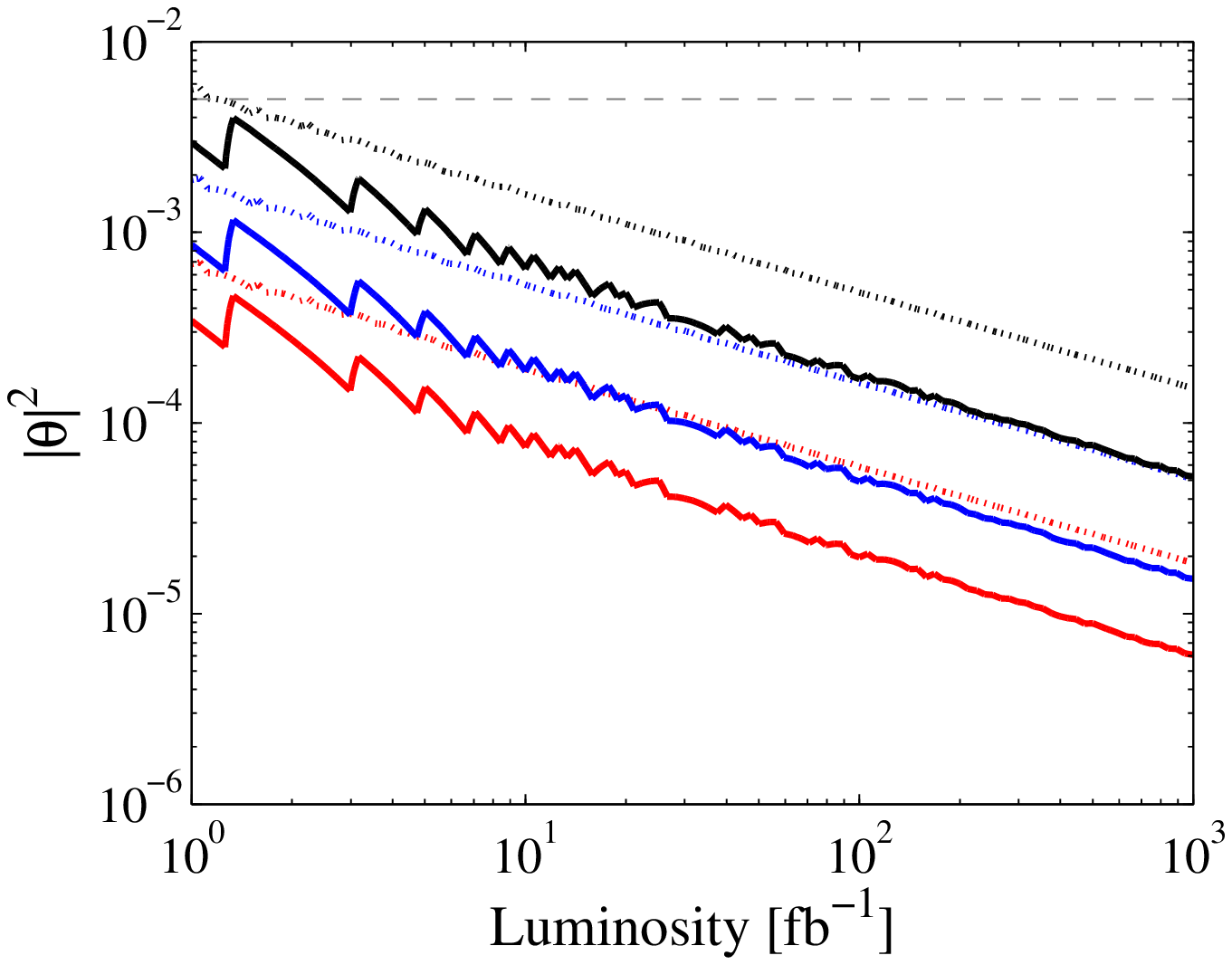} \\
\includegraphics[width=0.48\textwidth]{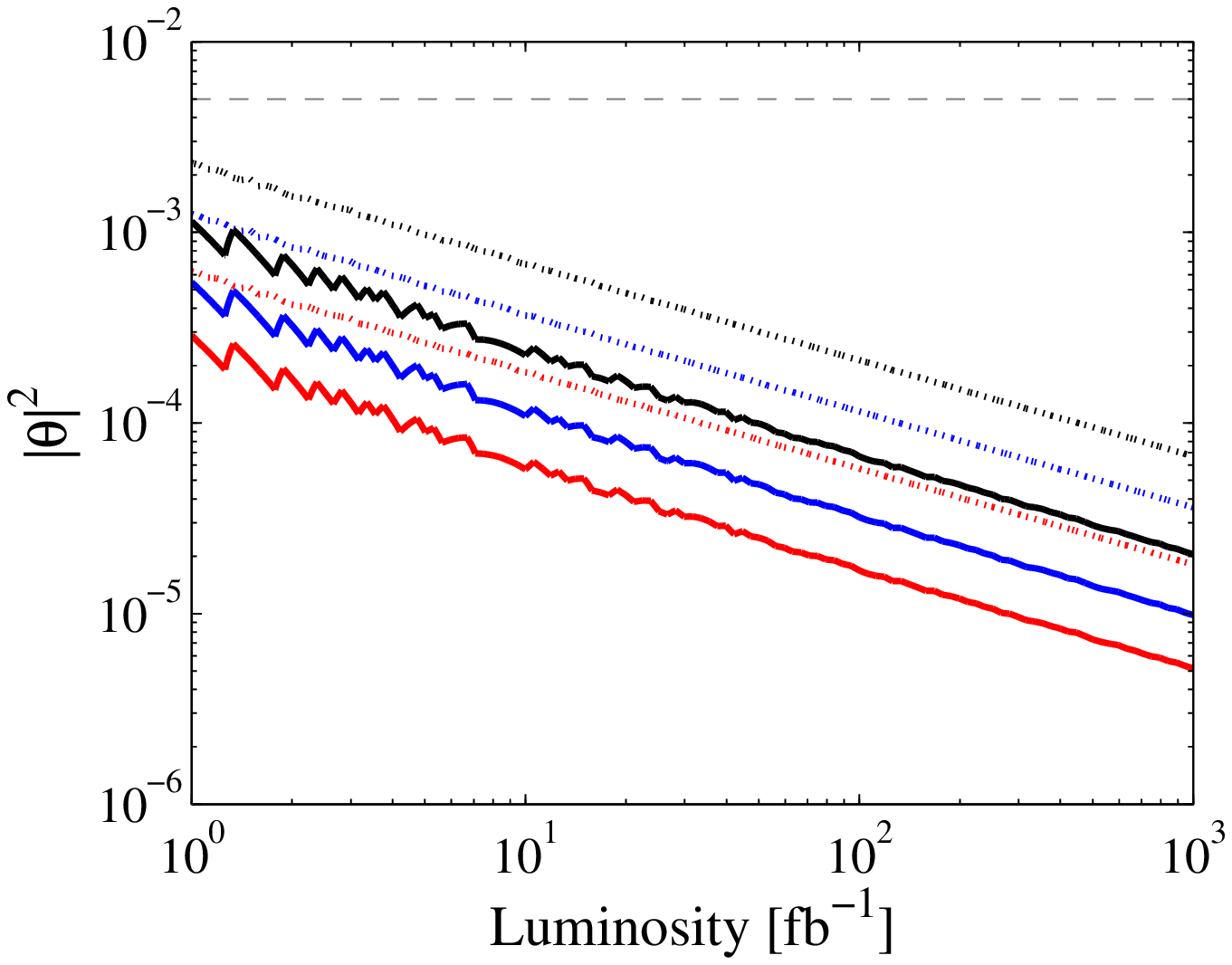}
\includegraphics[width=0.48\textwidth]{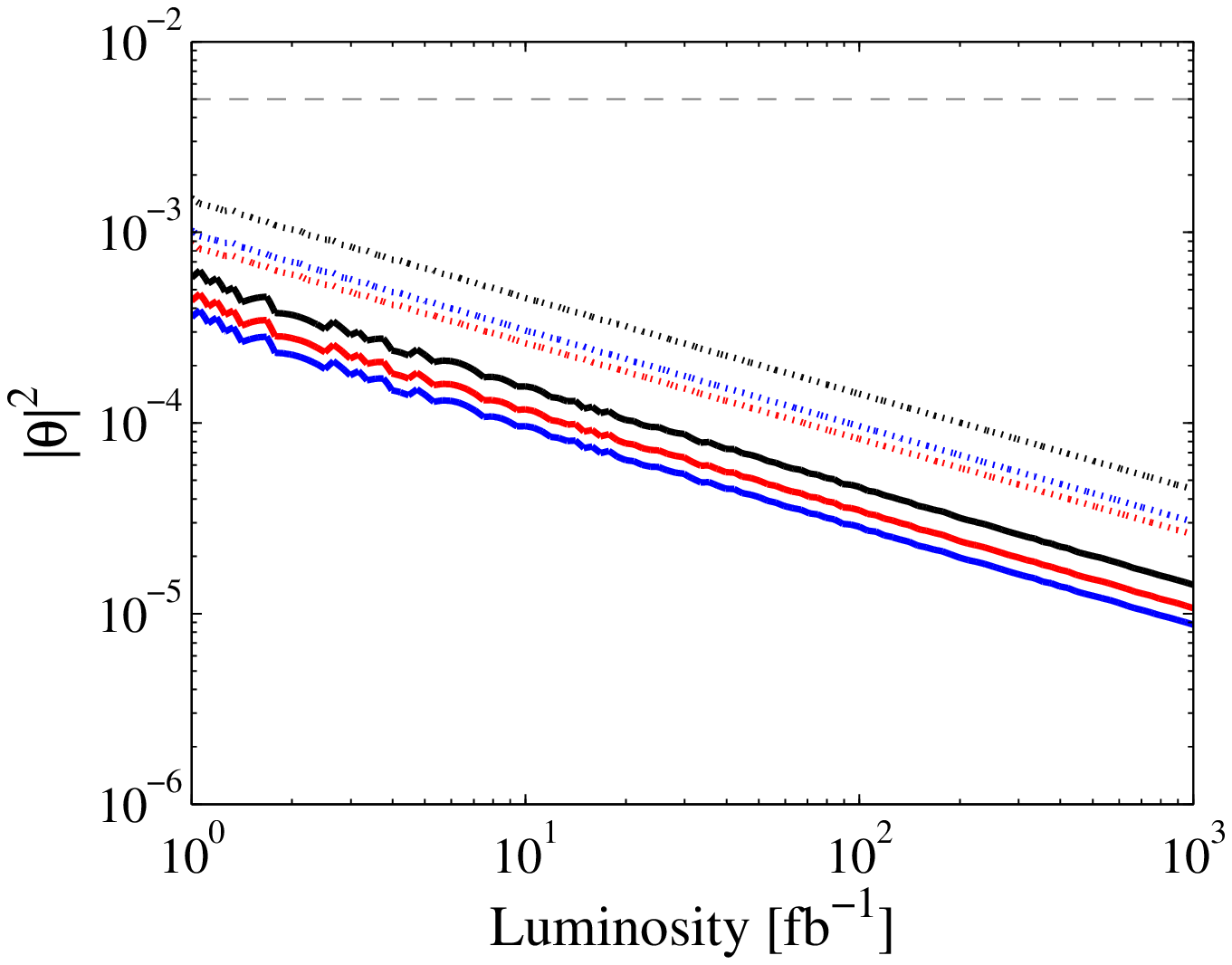}
\caption{The $90~\%$ CL sensitivity to the mixing between the heavy neutrino and electrons (upper panels) as a function of the luminosity at the 50~GeV (left) and 150~GeV (right) electron beam facilities. The lower panels are the corresponding plots for muons at the 500~GeV (left) and 2~TeV (right) muon beam facilities. Red, blue and black lines correspond to heavy neutrino masses of $M_N = 250$, 500 and 750~GeV, respectively. The solid/dotted lines are the sensitivities with/without cuts. }
\label{fig:sens}
\end{center}
\end{figure}

Finally, in \fig~\ref{fig:sens} we show an estimate of the achievable $90~\%$~CL sensitivity to the mixing between the heavy neutrino and the charged lepton as a function of the luminosity and for different values of the neutrino mass. The sensitivity has been defined as the minimum value of the mixing angle that can be excluded with at least 50~\% probability at the $90~\%$~CL in absence of signal. Given the low statistics expected for small mixings, a Poisson distribution was used to compute the sensitivity. The discrete nature of the Poisson distribution is the reason of the abrupt jumps in sensitivity depicted. The sensitivity estimate is conservative and based purely on a counting experiment without exploiting kinematic distributions beyond the $E_T$ and $p_T$ cuts such as the reconstructed mass $M_{jj\ell}$ of \fig~\ref{fig:Mnjje}. 
As can be seen from the figure, just based on a counting analysis sensitivities more than two orders of magnitude better than the current best fit from the invisible width of the $Z$ would be allowed in these facilities for luminosities of $100$ fb$^{-1}$.

\section{Summary and conclusions}
\label{sec:summary}

We have discussed the prospects of testing the existence of heavy Majorana neutrinos of a low scale type-I seesaw mechanism in lepton-proton collisions. 
The question of the Majorana nature of neutrinos together with the origin of neutrino masses is one of the most fundamental issues still unsolved in particle physics. The quest for answers to these unknowns is particularly challenging since all probes testing the Majorana character of neutrinos typically face a strong suppression of the signal due to the smallness of the active neutrino masses. 
In low scale seesaw mechanisms, however, this smallness originates from a cancellation between different large contributions that could, individually, lead to observable signals at searches such as \znbb{} decay or collider experiments. Thus, if the mass spectrum of the heavy Majorana neutrinos is such that some contributions are enhanced or suppressed by the kinematics of the process, the cancellation responsible for the smallness of neutrino masses will not take place.
This is particularly true in collider searches, as the one discussed here, given the fact that the contribution of a given neutrino can be enhanced via its s-channel production if the collider has enough energy to produce it on-shell. 

We have simulated the lepton number violating signal that the contribution of a heavy Majorana neutrino within the reach of the lepton-proton collider would provide. The simulations were performed for two types of prospective facilities. We have extended previous analyses of electron-proton colliders and updated them to the recent LHCeC proposals in which 7~TeV protons collide against electrons with energies between 50 and 150~GeV. As a more ambitious facility we also considered a muon-proton collider inspired by the efforts towards a muon collider. We simulated muons with energies between 0.5 and 2~TeV colliding against 7~TeV protons. Such a facility would not only allow to probe smaller mixings and higher neutrino masses through higher cross sections of the signal at higher energies, but also provide a complementary search channel, since it would probe areas of the parameter space completely independent to the electron-proton collisions or \znbb{} searches. 

In order to simulate the signal, we have set the mixing between the new heavy neutrinos and the charged leptons to the value required to accommodate the $2 \sigma$ deviation of the invisible width of the $Z$ with respect to the Standard Model prediction. Even this optimistic assumption leads to a small signal, as expected from the challenging task of testing Lepton number violation. We have shown that, despite the smallness of the expected cross section, the facilities under study would allow for an observable signal in areas of the parameter space still allowed by present data. Moreover, the signal from these facilities is particularly clean, with a very small Standard Model background to obscure it. The dominant background is originated from $W$ production, with its subsequent decay into an $\ell^+$ and the original $\ell^-$ missed by the detector. Nevertheless, we have studied how the signal over background ratio could be improved further through different kinematic cuts so as to allow the search of even smaller values of the mixing angle, provided sufficiently high collider luminosities. A conservative estimate of the sensitivity to the mixing between the heavy neutrinos and charged leptons yields an improvement of more than two orders of magnitude with respect to the present constraints for luminosities of $100$ fb$^{-1}$.

We conclude that lepton-proton colliders provide a particularly clean probe of the elusive Majorana nature of neutrinos and the type-I seesaw mechanism and would complement other ongoing searches by exploring different parts of the allowed parameter space.

\begin{acknowledgments}

We acknowledge illuminating discussions with Paolo Checchia, Andrea Donini, Vicent Mateu and Juan Terron. MB is supported by the European Union through the European Commission Marie Curie Actions Framework Programme 7 Intra-European Fellowship: Neutrino Evolution and.
PC acknowledges financial support from the Comunidad Aut\'onoma de Madrid, project HEPHACOS S2009/ESP-1473 and the Spanish Government under the Consolider-Ingenio 2010 programme CUP, ``Canfranc Underground Physics'', Project Number CSD00C-08-44022, and under project FPA2009-09017 (DGI del MCyT, Spain). PC would also like to thank the Institute for Particle Physics Phenomenology at Durham where part of this work was completed. MB, PC and EFM acknowledge financial support from the European Community under the European Comission Framework Programme 7,
Design Study: EUROnu, Project Number 212372 (the EU is not liable for
any use that may be made of the information contained herein).
\end{acknowledgments}

\end{document}